\documentclass[12pt,A4]{article}
\usepackage{epsfig}
\usepackage{array}
\usepackage{amsmath}
\usepackage{amssymb}

\newcommand{\m}{\mathbf}

\newcommand{\be}{\begin{equation}}
\newcommand{\ee}{\end{equation}}
\newcommand{\bea}{\begin{eqnarray}}
\newcommand{\eea}{\end{eqnarray}}

\title{COMMENT ON THE RELATION BETWEEN THE COLLINEAR TRIPARTITION AND THE COLLECTIVE MODEL OF THE ATOMIC NUCLEI}
\author{ F. F. Karpeshin \\ Mendeleev All-Russian Research
Institute of Metrology \\ 190005 Saint-Petersburg, Russia }

\begin{document}
\maketitle

\large

	In Ref. \cite{tash}  the authors consider dynamics of spontaneous fission of $^{252}$Cf into three fragments: Zn, Ca and Sn, the lightest fragment is in the middle, from the point of view of possible collinearity of tri-partition. The authors come to an evident conclusion that 1) if the fragments are on one line, the fission is collinear; 2) if the middle Ca fragment is shifted by 0.5 fm from the axis, that is the nascent fragments form a triangular-like configuration, or 3) if the middle fragment is on the axis, but has a transverse initial velocity, then the fragments separate non-collinearly.
Let us consider the second case, baring in mind that the third one is largely reduced to it. If all the nuclei are placed on one line, as in the first case, the triangle degenerates into the line. Note that the primordial for experiment  question:  which case will be realized, remains unanswered in the paper. 

	At the same time, quite an unambiguous answer is published in Refs. \cite{3fiz,3fyaf}. It is shown that a configuration with displaced mean fragment cannot be realized in the case of fission of spinless nucleus of $^{252}$Cf. This configuration is strictly forbidden by the symmetry principles which are in the basis of the collective Bohr's model \cite{BM}. Indeed, such a triangular shape, being not axially symmetric, only can reply to the projection of the angular momentum $I$ on the nuclear axis $K > 0$.  In more detail, $K$ is a good quantum number. In quantum mechanics, the body cannot rotate around the axis of symmetry, and thus the rotational momentum is perpendicular to the axis of symmetry, and its projection $K$ onto the axis of symmetry is zero. At the same time, a nonzero value of $K$ can also be observed in an axially symmetric system due to quasiparticle excitations unrelated to rotation. If the system is a bit non-axially symmetric, then basically it will also rotate around an axis perpendicular to the axis of symmetry, and it will only  twist around the axis of symmetry slightly. In quantum mechanics, this is reflected in the fact that the wave function of the fissile nucleus has the following form \cite{BM}:
 \be
 \Psi^I_M (\m r_i) = \sum_K a_K D^I_{MK}(\theta, \phi, \vartheta)\chi_K(\m r'_i)\,.     \label{D}
 \ee
$I$ in Eq. (\ref{D}) is the nuclear spin, $M$ --- its projection in the laboratory frame. Wigner's $D$ functions from the Euler angles $\theta, \phi, \vartheta$ define the orientation of the nucleus in space and determine the angular distribution of the fragments.  $\m r_i$ and $\m r'_i$ are the nuclear variables ({\it e. g.}, nucleon coordinates) in the laboratory system and intrinsic coordinate system, related to the  nucleus, respectively.  

	The $z'$ axis thus coincides with the fission axis. And let us direct  the $x'$ axis in the plane of symmetry of the fissile nucleus (that is in the plane of the triangular configuration). In terms of the Euler angles, the rotation from the laboratory system $\m r$ to the intrinsic system $\m r'$ can be performed in three steps. First two of them, the rotations by $\theta$ about the $z$ axis and by $\phi$ about the new axis $x'$, respectively,  impose the $z$ and $z'$ axes with each other \cite{BM}. After which, it remains the third rotation by $\vartheta$ about the new $z'$ axis, in order to impose the $x$ axis at the $x'$ one. In the case of a axially-symmetric nucleus, this third rotation evidently might not be  needed, as all the azimuthal angles are equivalent. The only way to combine this picture with Eq. (\ref{D})  is to put $K$ = 0. And {\it vice versa}: in the case of asymmetric --- triangular configuration of the fissile nucleus, rotation by the angle $\vartheta$ is essential. Correspondingly, this excludes values of $K$ = 0 in Eq. (\ref{D}), leaving only values $K>0$ as adoptable. 

	As $I\geq K$, non-zero $K$ values are only possible if fissile nucleus has a non-zero spin. This is not the case if spontaneous fission of  $^{252}$Cf is considered. Therefore, it is only the symmetric configuration ``three in line'' which survives fission. Such fission was figuratively called ``co-axial'' in Ref. \cite{3fyaf}. Furthermore, even such a ``co-axial'' initial configuration is not enough yet for the final collinearity. It can be destroyed during spreading of the fragments under action of the accelerating Coulomb force between them. This question was also studied in detail in Refs. \cite{3fiz,3fyaf}.  Perpendicular to the fission axis component of the initial velocity arises at the moment of scission due to big spins and  large relative angular momentum of the fragments. Big spins of the fragments are confirmed by experiment. Such spins can be formed  due to the wriggling vibrations of the fissile nucleus \cite{nix}. The orbital momentum arises for compensation of the total spin of the fragments. Allowance for the  arising initial transverse velocity  of the fragments results in the final divergence of the spreading fragments. If the fragments could move completely freely after scission, then with reasonable initial conditions, all the fragments would stay collinear on a rotating fission axis. The necessary condition for the fragments to remain on the axis is that both the transverse and longitudinal velocity components remain proportional to the distance from c. m.  When the Coulomb force is switched on, it changes only the longitudinal component. This violates the proportionality: the middle (ternary) fragment is pushed to the c. m. by the both outer fragments. This reflects, specifically, in its small final kinetic energy \cite{tash,3fiz,3fyaf}. In turn, the middle fragment itself pushes both the side fragments out, which also works as to violate the proportionality. 
As a result, as soon as the middle fragment descends from the axis, this immediately triggers the transverse component in the Coulomb repulsion between the fragments, which enhances the further destruction of collinearity. The final value   of the divergence between heavy and light fragments, however, does not exceed 2$^\circ$, which value is not resolved by the setup of Ref. \cite{piat}.  Basing on the above consideration, the collinear trajectory in the case of true ternary spontaneous fission of $^{252}$Cf  was concluded to be most probable.

\end{document}